\pdfoutput=1

\newcommand{\subparagraph}{}

\documentclass[10pt, conference, compsocconf, letterpaper]{IEEEtran}
\IEEEoverridecommandlockouts

\usepackage{graphicx}
\usepackage{caption}
\usepackage{graphicx}
\usepackage{xcolor}
\usepackage{mathtools}
\usepackage{braket}
\usepackage{listings}
\usepackage{cite}
\usepackage{titlesec}
\usepackage{booktabs}
\usepackage{array}
\usepackage{hyperref}
\usepackage{subcaption}
\newcolumntype{L}[1]{>{\raggedright\let\newline\\\arraybackslash\hspace{0pt}}m{#1}}

\usepackage{latexsym}
\usepackage{qcircuit}

\newcommand{\controlb}{*!<0em,.025em>-=-<.0em>{\Box}} 
\newcommand{\ctrlb}[1]{\controlb \qwx[#1] \qw} 

\lstdefinelanguage{qasm}
{
  morekeywords={
    qubits,
    name,
    map,
    mov,
    cnot, h, x, y, x, not, cx, cz, cr, crk, prep_x, prep_z, prep_y, rx, ry, rz, s, t, sdag, tdag, measure, display, measure_x, measure_z, measure_parity,version
  },
  sensitive=false, 
  morecomment=[l]{\#}, 
  morecomment=[s]{/*}{*/}
}

\usepackage{xcolor}

\colorlet{punct}{red!60!black}
\definecolor{background}{HTML}{EEEEEE}
\definecolor{delim}{RGB}{20,105,176}
\colorlet{numb}{magenta!60!black}

\lstdefinelanguage{json}{
    basicstyle=\normalfont\ttfamily,
    breaklines=true,
    basicstyle=\scriptsize\ttfamily, 
    captionpos=b, 
    extendedchars=true, 
    tabsize=2, 
    columns=fixed, 
    keepspaces=true, 
    showstringspaces=false, 
    breaklines=true, 
    frame=bt, 
    numbers=left, 
    numberstyle=\tiny\ttfamily, 
    numbersep=-5pt,
    literate=
     *{0}{{{\color{numb}0}}}{1}
      {1}{{{\color{numb}1}}}{1}
      {2}{{{\color{numb}2}}}{1}
      {3}{{{\color{numb}3}}}{1}
      {4}{{{\color{numb}4}}}{1}
      {5}{{{\color{numb}5}}}{1}
      {6}{{{\color{numb}6}}}{1}
      {7}{{{\color{numb}7}}}{1}
      {8}{{{\color{numb}8}}}{1}
      {9}{{{\color{numb}9}}}{1}
      {:}{{{\color{punct}{:}}}}{1}
      {,}{{{\color{punct}{,}}}}{1}
      {\{}{{{\color{delim}{\{}}}}{1}
      {\}}{{{\color{delim}{\}}}}}{1}
      {[}{{{\color{delim}{[}}}}{1}
      {]}{{{\color{delim}{]}}}}{1}
}

\definecolor{eclipseBlue}{RGB}{42,0.0,255}
\definecolor{eclipseGreen}{RGB}{63,127,95}
\definecolor{eclipsePurple}{RGB}{127,0,85}

\def\note#1{}
\def\note#1{\textbf{\color{red}[#1]}}

\newcommand\Mark[1]{\textsuperscript{#1}}

\begin{document}

\title{OpenQL : A Portable Quantum Programming Framework for Quantum Accelerators \\[1.25ex]
}

\author{\normalfont\large N. Khammassi\Mark{\S}~\Mark{\ddag}~\Mark{*}\thanks{\textsuperscript{* } Currently Affiliated to Intel Labs, Intel Corporation, Oregon, USA}, I. Ashraf\Mark{\S}~\Mark{\ddag}, J. v. Someren\Mark{\S}~\Mark{\ddag},  R. Nane\Mark{\S}~\Mark{\ddag}, A. M. Krol\Mark{\S}~\Mark{\ddag}, M.A. Rol\Mark{\P}~\Mark{\ddag}, \\ L. Lao\Mark{\S}~\Mark{\ddag}~\Mark{**}\thanks{\textsuperscript{**} Currently Affiliated to Department of Physics and Astronomy, University College London, UK}, K. Bertels\Mark{\S}, C. G. Almudever\Mark{\S}~\Mark{\ddag} \\
    \newline \\ \vspace{0.2cm}
    \Mark{\S} Quantum \& Computer Engineering Dept., Delft University of Technology Delft, The Netherlands \\
    \vspace{0.2cm}
    \Mark{\ddag} QuTech, Delft University of Technology, The Netherlands \\ \vspace{0.2cm}
    \Mark{\P} Kavli Institute of Nanoscience, Delft University of Technology, The Netherlands}

\maketitle

\begin{abstract}

With the potential of quantum algorithms to solve intractable classical problems, quantum computing is rapidly evolving and more algorithms are being developed and optimized. Expressing these quantum algorithms using a high-level language and making them executable on a quantum processor while abstracting away hardware details is a challenging task. Firstly, a quantum programming language should provide an intuitive programming interface to describe those algorithms. Then a compiler has to transform the program into a quantum circuit, optimize it and map it to the target quantum processor respecting the hardware constraints such as the supported quantum operations, the qubit connectivity, and the control electronics limitations. In this paper, we propose a quantum programming framework named OpenQL, which includes a high-level quantum programming language and its associated quantum compiler. We present the programming interface of OpenQL, we describe the different layers of the compiler and how we can provide portability over different qubit technologies. Our experiments show that OpenQL allows the execution of the same high-level algorithm on two different qubit technologies, namely superconducting qubits and Si-Spin qubits. Besides the executable code, OpenQL also produces an intermediate quantum assembly code (cQASM), which is technology-independent and can be simulated using the QX simulator.  
 
\end{abstract}

\textit{Quantum Compiler, Quantum Computing, Quantum Circuit, Quantum Processor.}


\section{Introduction}

Since the early formulation of the foundations of quantum computing, several quantum algorithms have been designed for solving intractable classical problems in different application domains. For instance, the introduction of Shor's algorithm~\cite{Shor:1997} outlined the significant potential of quantum computing in speeding up prime factorization. Later, Grover's search algorithm~\cite{Grover1997} demonstrated quadratic speedup over its classical implementation counterpart. The discovery of these algorithms boosted the development of different physical qubit implementations such as superconducting qubits~\cite{versluis2017scalable}, trapped ions~\cite{ion-trap-qprocessor} and semiconducting qubits~\cite{watson2018}. 

In the absence of a fully programmable quantum computer, the implementation of these algorithms on real quantum processors is a tedious task for the algorithm designer, especially in the absence of deep expertise in qubit control electronics. 
In order to make a quantum computer programmable and more accessible to quantum algorithm designers similarly to classical computers, several software and hardware layers are required \cite{almudever2017engineering}: at the highest level, an intuitive quantum programming language is needed to allow the programmer to express the quantum algorithm without worrying about the hardware details. Then, a compiler transforms the algorithm into a quantum circuit and maps and optimizes it for a given quantum processor. Ultimately, the compiler produces an executable code which can be executed on the target micro-architecture controlling the qubits. 
A modular quantum compiler would ideally not expose low-level hardware details and its constraints to the programmer to allow portability of the algorithm over a wide range of quantum processors and qubit technologies. \\


In this paper we introduce OpenQL\footnote{OpenQL documentation: https://openql.readthedocs.io}, an open-source\footnote{OpenQL source code: https://github.com/QE-Lab/OpenQL} high-level quantum programming framework. OpenQL is mainly composed of a quantum programming interface for implementing quantum algorithms independently from the target platform, and a compiler which can compile the algorithm into executable code for various target platforms and qubit technologies such as superconducting qubits and semiconducting qubits.

The rest of the paper is organized as follows. Section~\ref{sec:related-work} provides a brief account of the related work. The necessary background for the quantum accelerator model is given in Section~\ref{sec:accelerator-model}. OpenQL architecture is detailed in Section~\ref{sec:openql-arch}, followed by the discussion of quantum programming interface provided by OpenQL in Section~\ref{sec:prog-interface}. OpenQL compilation passes are presented in Section~\ref{sec:compilation-stages}, where it is shown how the quantum code is decomposed, optimized, scheduled, and mapped on the target platform. Some of the works in which we utilized OpenQL to compile quantum algorithms on different quantum processors using different qubit technologies, are briefly mentioned in Section~\ref{sec:results}. Finally, Section~\ref{sec:conclusion} concludes the paper.

\section{Related Work}
\label{sec:related-work}

Some of the initial work in the field of quantum compilation has been theoretical \cite{Bettelli2003, omer98aprocedural, selinger_2004, lambda-calculus-2005, zorzi_2016, qwire-2017}. Now that quantum computers are a reality, various compilation and simulation software frameworks have been developed. A list of open-source compilation projects is available at~\cite{qcompiler_list}, and a list of quantum simulators is available at~\cite{qsim_list}. In the following, we provide a brief list of recent active works in the field of quantum compilation in chronological order. The reader is referred to a recent overview and comparison of gate-level quantum software platforms~\cite{qsoft-overview_2019}.

\begin{itemize}

\item ScaffCC has been presented as a scalable compilation and analysis tool for quantum programs~\cite{scaffcc2014, ScaffCC15}. It is based on LLVM compilation framework. ScaffCC compiles Scaffold language~\cite{scaffold-2012}, which is a pure quantum language embedded into the classical \emph{C} language.


\item Microsoft proposed a domain-specific language Q\#\cite{q-sharp-18} and Quantum Development Kit (QDK) to compile and simulate quantum programs. At the moment, QDK does not target a real quantum computer, however, programs can be executed on the provided software backend. 

\item ProjectQ~\cite{projectq-18} is an open-source software framework that allows the expression of a quantum program targeting IBM backend computers as well as simulators. ProjectQ allows programmers to express their programs in a language embedded in python. Apart from low-level gate description, meta-instructions are provided to add conditional control, compute, un-compute, and repeating sections of code a certain number of times.

\item IBM's Qiskit \cite{Qiskit-19} is an open-source quantum software framework that allows users to express their programs in python and compiles them to OpenQASM targeting the IBM Q Experience\cite{IBMQE}. Qiskit allows users to explicitly allocate quantum and classical registers. Quantum operations are performed on quantum registers, and after measurement, classical results are stored in classical registers.

\item Quilc~\cite{quilc-20} is an open-source quantum compiler for compiling Rigetti's Quil language~\cite{pyquil-16}. The focus of the authors is on the noisy intermediate scale quantum programs, allowing the programmers to compile quantum programs to byte code, which can be interpreted by control electronics. This allows programmers to execute programs not only on a software simulator but also on real quantum processor.
\end{itemize}

OpenQL has some common characteristics with the compilers above, such as being an open-source, modular quantum compilation framework that is capable of targeting different hardware backends. However, the distinctive and, at the same time, the primary motivation behind OpenQL is that it is a generic and flexible compiler framework. These requirements directly translated into the OpenQL design to support multiple \textit{configurable} backends through its platform configuration file (Section~\ref{sec:qplatform}). Finally, OpenQL is one of the engines behind QuTech's Quantum Inspire~\cite{misc:qinspire} platform, where the user can gain access to various technologies to perform quantum experiments enabled through the use of OpenQL's plugin-able backends and its ability to generate executable code (Section~\ref{sec:eqasm}).


\section{Quantum Accelerator Model}
\label{sec:accelerator-model}


Accelerators are used in classical computers to speed up specific types of computation that can take advantage of the execution capabilities of the accelerator such as massive parallelism, vectorization or fast digital signal processing... OpenQL adopts this heterogeneous computing model while using the quantum processor as an accelerator and provides a programming interface for implementing quantum algorithms involving both classical computation and quantum computation. 

\begin{figure*}[t]
\centering
\includegraphics[scale=0.4]{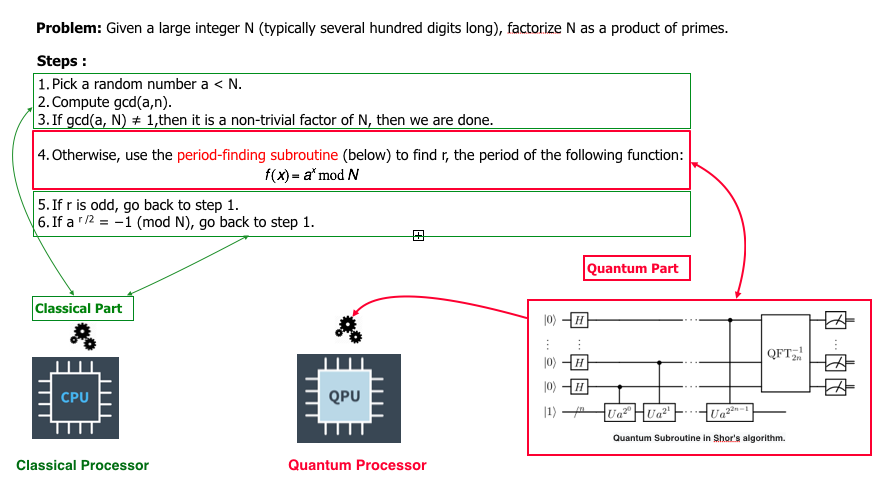}
\caption{Shor's algorithm is composed of both classical computations and quantum computations.}
\label{fig:shor}
\end{figure*}

\subsection{Heterogeneous Computing}



Heterogeneous computing~\cite{het-comp, hsa} is a computing model where a program is executed jointly on a general-purpose processor or host processor and an accelerator or co-processor. The general-purpose processor is capable of executing not only general computations such as arithmetic, logic or floating point operations, but also controlling various accelerators or co-processors. The accelerators or co-processors are specialized processors designed to accelerate specific types of computation such as graphic processing, digital signal processing and other workloads that can take advantage of vectorization or massive thread-level parallelism. Therefore the accelerator can speedup a part of the computation traditionally executed on a general purpose processor. The computation is then offloaded to the accelerator to speed up the overall execution of the target program. Examples of accelerators are the Intel Xeon Phi co-processor \cite{XeonPhi}, Digital Signal Processors (DSP) ~\cite{omap3530}, Field Programmable Gate Array (FPGA)~\cite{zynq, vassiliadis2004} that can be also utilized as accelerators to parallelize computations and speed up their execution. Finally General-Purpose Computation on Graphics Processing Units (GPGPU) uses GPU as accelerator ~\cite{gpgpu} to speed up certain types of computations.

\subsection{Quantum Processors as Accelerators}


The OpenQL programming framework follows a heterogeneous programming model which aims to use the quantum processor as a co-processor to accelerate the part of the computation which can benefit from the quantum speedup. A quantum algorithm is generally composed of classical and quantum computations. For instance Shor's algorithm is a famous quantum algorithm for prime number factoring; as shown in Figure~\ref{fig:shor} the algorithm includes classical computations such as the Greatest Common Divisor (GCD) computation which can be executed efficiently in a traditional processor, and a quantum part such as the Quantum Fourier Transform which should be executed on a quantum processor.

OpenQL uses traditional host languages, namely C++ and Python, to define a programming interface which allows the expression of the quantum computation and the communication with the quantum accelerator: the quantum operations are executed on the quantum processor using a dedicated micro-architecture and the measurement results are collected and sent back to the host program running on the classical processor. While non time-critical classical operations can be executed on the host processor, time-critical classical operations that need to be executed within the coherence time of the qubits, such as in error correction quantum circuits, can be offloaded to the accelerator to provide fast reaction time and avoid communication overhead between the host PC and the accelerator.



\section{OpenQL Architecture}
\label{sec:openql-arch}


\begin{figure*}[th]
\centering
\includegraphics[scale=0.4]{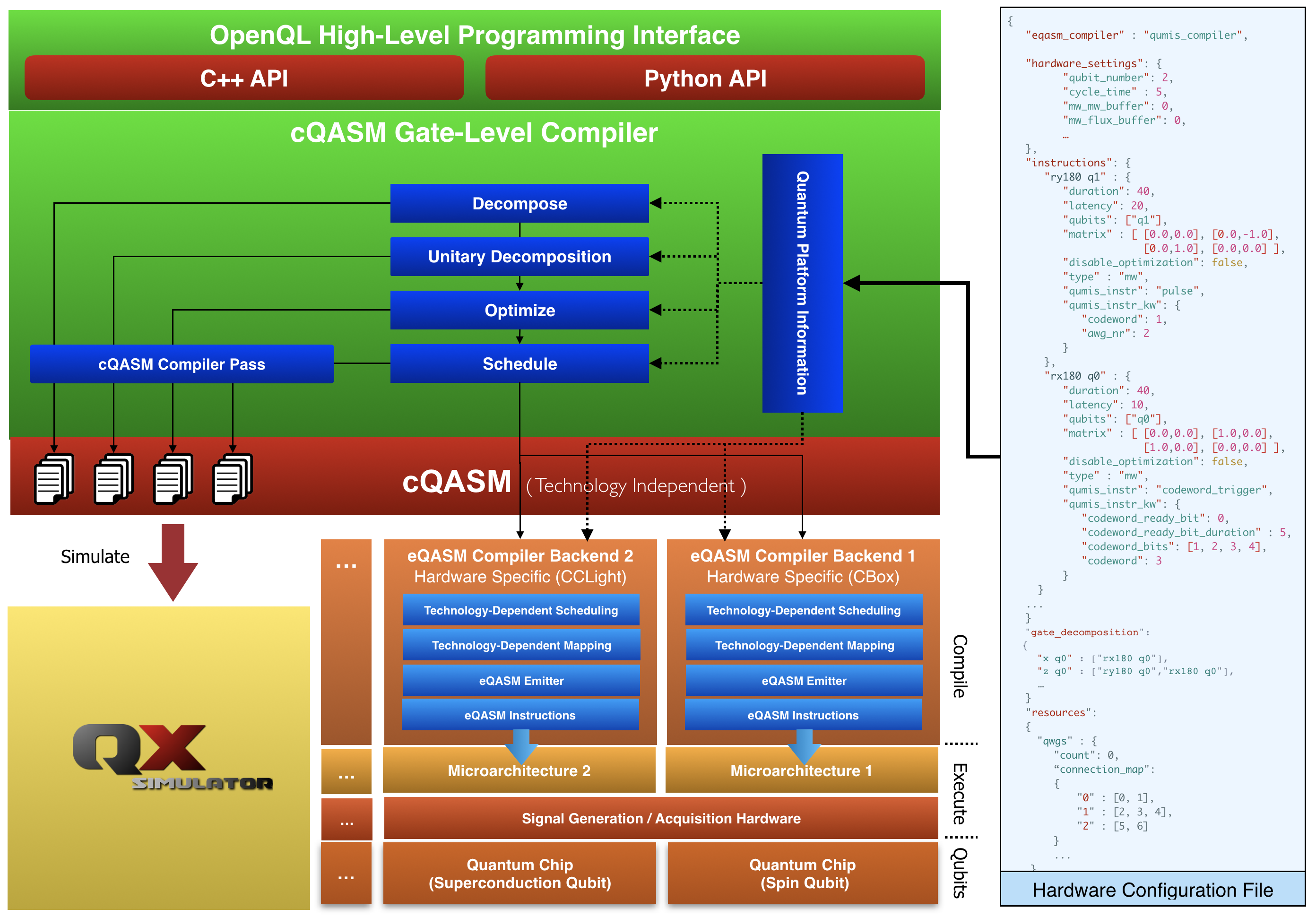}
\caption{OpenQL Compiler Architecture}
\label{fig:compiler_infrastructure}
\end{figure*}


Figure~\ref{fig:compiler_infrastructure} depicts OpenQL framework which exposes a high-level programming interface to the user at the top. The compiler implements a layered architecture which is composed mainly of two parts: a set of hardware-agnostic compilation passes that operate at the quantum gate level, and a set of low-level technology-specific backends which can target different quantum processors with specific control hardware. The goal of those backends is to enable compiling the same quantum algorithm for a specific qubit technology without any change in the high-level code and making the hardware details transparent to the programmer. Moreover, this architecture allows the implementation of new backends to extend the support to other qubit technologies and new control hardware whenever needed. As the qubit control hardware is constantly evolving in the last years, this flexibility and portability over a wide range of hardware is crucial. This enhances the productivity and ensures the continuity of the research efforts towards a full-stack quantum computer integration. 

The Quantum Assembly Language (QASM) is the intermediate layer which draws the abstraction line between the high-level hardware-agnostic layers (gate-level compilation stages) and the low-level hardware-specific layers. The low-level layers are implemented inside a set of interchangeable backends each targeting a different microarchitecture and/or a different qubit technology. \\ 

The OpenQL framework is composed mainly of the following layers:
\begin{itemize}
    \item A High-level programming interface using a standard host language namely C++ or Python to express the target quantum algorithm as a quantum program. 
    \item A quantum gate-level compiler that transforms the quantum program into a quantum circuit, optimizes it, schedules it and maps it to the target quantum processor to comply to the different hardware constraints such as the limited qubit connectivity. 
    \item The last stage of the gate-level compilation produces a technology-independent Common Quantum Assembly code (cQASM) \cite{khammassi2018cqasm} which describes the final quantum circuit while abstracting away the low-level hardware details such as the target instruction set architecture, or the quantum gate implementation which differ across the different qubit technologies 
    For now, our compiler targets Superconducting qubits and Si-Spin qubits but can be easily extended to other qubit technologies. 
    The produced QASM code complies with the Common QASM 1.0 syntax and can be simulated in our QX simulator~\cite{Khammassi2017} to debug the quantum algorithm and  evaluate its performance for different quantum error rates.
    \item At the lowest level, different eQASM~\cite{eQASM} ({\it executable QASM}) backends can be used to compile the QASM code into instructions which can be executed on a specific micro-architecture, e.g. the QuMA micro-architecture described in \cite{Fu2017}. At this compilation level, very detailed information about the target hardware setup, stored in a hardware configuration file, is used to generate an executable code which takes into account various hardware details such as the implementation of the quantum gates, the connectivity between the qubits and the control instruments, the hardware resource dependencies, the quantum operation latencies and the operational constraints.
\end{itemize} 

\section{Quantum Programming Interface}
\label{sec:prog-interface}

OpenQL provides three main interfaces to the developer, namely Quantum Kernel, Quantum Program and Quantum Platform.

\subsection{Quantum Kernel}
A \textit{Quantum Kernel} is a quantum functional block which consists of a set of quantum or classical instructions and performs a specific quantum operation. For instance, the kernel could be dedicated to creating a bell pair while another could be dedicated to teleportation or decoding. In OpenQL a \textit{Quantum Kernel} can be created as shown in Code Example 1 where three kernels are created: i) the "init" kernel for initializing the qubits, ii) the "epr" kernel to create a Bell pair, iii) the "measure" kernel to measure the qubits. These kernels are then added to the main program, and compiled while enabling the compiler optimizations and the As Late As Possible (ALAP) scheduling scheme.
In code example 2, the same code is written in the C++ programming language. Note that the programming API of C++ is identical to the Python API.\\ 


\begin{lstlisting}[language=Python,caption={OpenQL Python code creating a Bell pair}, label={lst:py_bell}, ,captionpos=b]
   import openql as ql
   
   # load the hardware config of the target platform   
   transmon = ql.quantum_platform('transmon', “hardware_config.json”);

   # we create the main quantum program
   prog = program('bell_pair',2,transmon)

   # create new kernels
   k1 = kernel('init'); # prepare q0 and q1 in zero state
   k1.prepz(0);     
   k1.prepz(1);  
   k2 = kernel('epr'); # create a bell pair
   k2.hadamard(0);  # H q0
   k2.cnot(0,1);    # CNOT q0,q1
   k3 = kernel('measure'); # measure
   k3.measure(0);   
   k3.measure(1);   
   # add kernel to the quantum program
   prog.add_kernel(k1);    
   prog.add_kernel(k2);
   prog.add_kernel(k3);
   
   // compile and optimize the program
   prog.compile(optimize=true,schedule='ALAP');
\end{lstlisting}

\begin{lstlisting}[language=C++,caption={OpenQL C++ code creating a Bell pair},captionpos=b]
   #include <ql/openql.h>
   
   // load the hardware config of the target platform   
   ql::quantum_platform transmon(“transmon”, “hardware_config.json”);

   // create quantum program
   ql::program prog(“prog”,2,transmon);

   // create new kernels
   ql::quantum_kernel k1("init"); // prepare q0 and q1 in zero state
   k1.prepz(0);     
   k1.prepz(1);  
   ql::quantum_kernel k2("epr"); // create a bell pair
   k2.hadamard(0);  // H q0
   k2.cnot(0,1);    // CNOT q0,q1
   ql::quantum_kernel k3("measure"); // measure
   k3.measure(0);   
   k3.measure(1);   
   // add kernels to the quantum program
   prog.add(k1);    
   prog.add(k2);
   prog.add(k3);
   
   // compile and optimize the program
   prog.compile(optimize=true,schedule="ALAP");
\end{lstlisting}

\begin{table}[h!]
\centering
\caption{Supported Quantum Gates} 
\label{tab:gates}
\begin{tabular}{|L{1.6cm} | L{3.4cm} | L{2.1cm} |}
\hline
\textbf{Quantum Gate}  & \textbf{Description} & \textbf{Example}  \\ \hline
I & Identity  &  kernel.identity(3)             \\ \hline
H & Hadamard & kernel.hadamard(0)             \\ \hline
X & Pauli-X & kernel.x(1)             \\ \hline
Y & Pauli-Y & kernel.y(3)           \\ \hline
Z & Pauli-Z & kernel.z(7)           \\ \hline
Rx & Arbitrary x-rotation & kernel.rx(0, 3.14)      \\ \hline
Ry & Arbitrary y-rotation & kernel.ry(5, 1.75)      \\ \hline
Rz & Arbitrary z-rotation & kernel.rz(2, 0.5)      \\ \hline
X90 & R\_{x}($\pi/2$) & kernel.x90(7)           \\ \hline
Y90 & R\_{y}($-\pi/2$) & kernel.y90(5)          \\ \hline
mX90 & R\_{x}($-\pi/2$) & kernel.mx90(2)          \\ \hline
mY90 & R\_{y}($-\pi/2$) & kernel.my90(1)          \\ \hline
S & Phase & kernel.s(3)             \\ \hline
Sdag    & Phase dagger &  kernel.sdag(13)          \\ \hline
T & T & kernel.t(2)             \\ \hline
Tdag & T dagger &  kernel.tdag(12)          \\ \hline
CNOT & CNOT & kernel.cnot(3,5)       \\ \hline
Toffoli & Toffoli & kernel.toffoli(3,5,7) \\ \hline
CZ & CPHASE & kernel.cz(1,2)         \\ \hline
SWAP & Swap & kernel.swap(0,3)       \\ \hline \\ \hline
Custom & Custom gate & kernel.gate("name",2) \\ \hline 
\end{tabular}
\end{table}

OpenQL supports standard quantum operations as listed in Table~\ref{tab:gates}. To allow for further flexibility in implementing the quantum algorithms, custom operations can also be defined in a hardware configuration file. These operations can either be independent physical quantum operations supported by the target hardware or a composition of a set of physical operations. Once defined in the configuration file of the platform, the new operation can be used in composing a kernel as any other predefined standard operation. This allows for more flexibility when designing a quantum algorithm or a standard experiment used for calibration or other purposes.


\subsection{Quantum Program}
As the quantum kernels implement functional blocks of a given quantum algorithm, a "\textit{quantum\_program}" is the container holding those quantum kernels and implementing the complete quantum algorithm. For instance, if our target algorithm is a quantum error correction circuit which includes the encoding of the logical qubit, the error syndrome measurement, the error correction and finally the decoding, we can create four distinct kernels which implement these four blocks, and we can add these kernels to our program. The program can then be  compiled and executed on the target platform.

\subsection{Quantum Platform}
\label{sec:qplatform}
A "\textit{quantum\_platform}" is a specification of the target hardware setup including the quantum processor and its control electronics. The specification includes the description of the supported quantum operations and their attributes such as the duration, the built-in latency of each operation and the mathematical description of the supported quantum operation such as its associated unitary matrix.

\section{Quantum Gate-Level Compilation}
\label{sec:compilation-stages}


The first compilation stages of OpenQL are performed at the quantum gate-level while abstracting the low-level hardware implementation on the target device as much as possible. The high-level compilation stages include the decomposition of the quantum operations, the optimization and the scheduling of the decomposed quantum circuit. The gate-level compilation layers can produce a technology-agnostic quantum assembly code called common QASM (cQASM) that can be simulated using the QX Simulator \cite{khammassi2017qx}.

\subsection{Gate Decomposition}

OpenQL supports decomposition of multi-qubit gates to 1 and 2 qubit gates, as well as control decomposition of multiple gates which are controlled by 1 or more qubits. Gates which are expressed as unitary matrices can also be decomposed to rotation and controlled-not gates.


\subsubsection{Multi-qubit Gate Decomposition}

\begin{figure}[h!]
\centering
\includegraphics[width=0.8\linewidth]{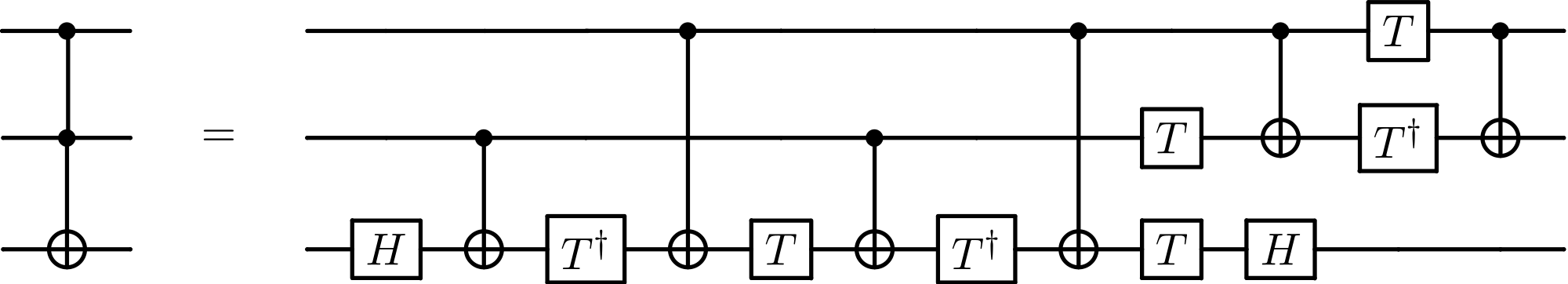}
\caption{Toffoli Gate Decomposition}
\label{fig:toffoli_decompose}
\end{figure}

In the first step, quantum gates are decomposed into a set of elementary operations from a universal gate set. For instance, as shown in Fig. \ref{fig:toffoli_decompose}, the Toffoli gate can be decomposed into a set of single and two-qubit gates using different schemes such as in \cite{NielsenChuangBook} or \cite{Amy2013}.

The decomposition of gates with more than two qubit operands is necessary to enable the later mapping stage which can only deal with available single and two-qubit gates that are available on the target physical implementation.
Furthermore, this decomposition allows us to perform fine-grain optimization through fusing operations and extracting parallelism using gate dependency analysis. When a physical target platform and its supported physical operations are specified in the configuration file, by doing this decomposition the compiler makes sure that the remaining operations are the target primitive operations that are supported by the target platform. The hardware configuration specification is detailed in Section~\ref{subsec:hw_config}. We note that we can disable this decomposition stage when the QX simulator backend \cite{Khammassi2017} is targeted as QX can simulate composite gates such as the Toffoli gate or arbitrary controlled rotations that are not necessarily available for many physical devices. \\

\begin{figure}[h!]
\centering
\includegraphics[width=0.75\linewidth]{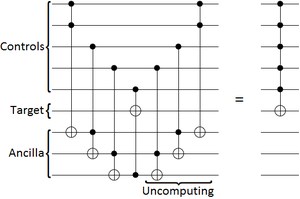}
\caption{Multi-qubit Controlled Decomposition}
\label{fig:multi-control-decomposition}
\end{figure}

Multi-qubit controlled gates can also be decomposed to 2-qubit controlled gates as discussed in~\cite{NielsenChuangBook} based on the scheme shown in Figure~\ref{fig:multi-control-decomposition}.

\begin{lstlisting}[language=Python,caption={OpenQL Multi-qubit Controlled kernel}, label={lst:mqc-kernel}, captionpos=b]
    ...
    k.gate("x", [0])
    k.gate("y", [0])
    k.gate("h", [0])
    ...
    
    # generate controlled version of k.
    # qubit 1 is used as control qubit
    # qubit 2 is used as ancilla qubit
    ck.controlled(k, [1], [2])
\end{lstlisting}

OpenQL further extends the facility of control decomposition to multiple gates (kernel). This is achieved by generating the controlled version of a kernel by using the $controlled()$ API as depicted in Code example ~\ref{lst:mqc-kernel} and then applying decomposition.

\subsubsection{Unitary Gate Decomposition}

It has been demonstrated that a universal quantum computer can simulate any Turing machine~\cite{deutsch1985quantum} and any local quantum system~\cite{lloyd1996universal}. A set of gates is called universal if they can be used to constitute a quantum circuit that can approximate any unitary operation to arbitrary accuracy.  

\begin{equation}
\label{eqn:1}
\begin{split}
H = \frac{1}{\sqrt{2}} \begin{bmatrix} 1  & 1 \\ 1 & -1 \end{bmatrix} 
\end{split}
\quad
\begin{split}
T = \begin{bmatrix} 1  & 0 \\ 0 & e^{i\pi/4} \end{bmatrix}
\end{split}
\end{equation}

\begin{equation}
\label{eqn:2}
\begin{split}
 X = \begin{bmatrix} 0  & 1 \\ 1 & 0 \end{bmatrix} 
\end{split}
\quad
\begin{split}
 Y = \begin{bmatrix} 0  & -i \\ i & 0 \end{bmatrix}
\end{split}
\quad
\begin{split}
 Z = \begin{bmatrix} 1  & 0 \\ 0 & -1 \end{bmatrix}
\end{split}
\end{equation}

\begin{equation}
\label{eqn:cnot}
CNOT = \begin{bmatrix} 1  & 0 & 0 & 0 \\ 0  & 1 & 0 & 0 \\ 0  & 0 & 0 & 1 \\ 0  & 0 & 1 & 0 \end{bmatrix}
\end{equation}

It has been proven that any unitary operation can be approximated to arbitrary accuracy by using only single qubit gates such as given in equations \ref{eqn:1} and \ref{eqn:2} and the CNOT gate, as given in equation~\ref{eqn:cnot}~\cite{NielsenChuangBook}.

A unitary matrix is used to represent each quantum operation of our quantum circuit to enable decomposition and fusing of quantum operations. The unitary matrix representation of gates is a useful mathematical tool which allows the compiler to efficiently fuse quantum operations using simple matrix multiplications and Kronecker product computations. Combining quantum gates is particularly useful for reducing the number of quantum operations and thus the overall execution time of a quantum algorithm to perform the largest possible number of quantum operations within the coherence time of the qubits. For instance, combining a set of single qubit rotations can be cancelled out if their fusion is equivalent to an identity operation which can be removed from the quantum circuit.  \\

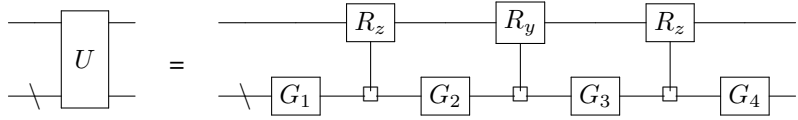
\begin{figure*}[h]
\centering
\leavevmode
\Qcircuit @C=1em @R=.7em {
&\qw &\multigate{1}{U}   & \qw   & 
& &\qw &\qw & \gate{R_z} &\qw & \gate{R_y} &\qw & \gate{R_z} & \qw &\qw\\
&\backslash\qw   &\ghost{U}          &\qw    &\push{\rule{.3em}{0em}\raisebox{2.2em}{=}\rule{.3em}{0em}}
& &\backslash\qw &\gate{G_1} & \ctrlb{-1} &\gate{G_2} & \ctrlb{-1} &\gate{G_3} & \ctrlb{-1} & \gate{G_4}&\qw }
\caption{Quantum Shannon Decomposition\cite{synthesisofquantumlogiccircuits}}
\label{circ:quantumshannon}
\end{figure*}

Any quantum gate can be fully specified using a unitary matrix, and any unitary matrix can be decomposed into a finite number of gates from some universal set. In OpenQL, this is achieved using Quantum Shannon Decomposition\cite{synthesisofquantumlogiccircuits} as show in Figure \ref{circ:quantumshannon}, which has been implemented using the C++ Eigen library~\cite{misc:eigendoc}. The universal set of gates used are the arbitrary y-rotation, the arbitrary z-rotation and the controlled-not gate. The matrices for these are shown in equations~\ref{eqn:cnot} and \ref{eqn:rotyz}.

\begin{equation}
\label{eqn:rotyz}
\begin{split}
R_y(\theta) = \begin{bmatrix} cos\theta /2 & sin\theta /2 \\ -sin \theta /2 & cos \theta / 2\end{bmatrix}
\end{split}
\quad
\begin{split}
R_z(\theta)=\begin{bmatrix} e^{-i\theta /2} & 0\\ 0 & e^{i\theta/2}\end{bmatrix}\end{split}
\end{equation}

At each level of the recursion, a unitary gate $U$ is decomposed into four unitary gates spanning one less qubit, and three uniformly controlled rotation gates. The latter are decomposed using the technique from \cite{art:quantumcircuitsforgeneralmultiqubit}, and the algorithm is called again on the smaller unitary gates. This recursion continues until the one-qubit unitary gates can be implemented using ZYZ-decomposition \cite{barenco1995elementary}.

For an n-qubit unitary, the decomposition results in $U(n) = 3/2*4^n-3/2*2^n$ rotation gates and $C(n) = 3/4*4^n-3/2*2^n$ controlled-not gates. These gates are added to the circuit and passed on to the next stages in the compilation.

\subsection{Gate-Level Optimization}

\subsubsection{Gate Dependency Analysis}
Once the quantum operations have been decomposed into a sequence of elementary operations, the gate dependency is analyzed and represented in the form of a Direct Acyclic Graph (DAG) where the nodes represent the quantum gates and the edges the dependency between them. We refer to this graph as the Gate Dependency Graph (GDG). Beside extracting the parallelism from the quantum circuit, the GDG allows reordering the gates with respect to their dependencies and helps  extracting local gate sequences that can potentially fused into smaller sequence of operations or even cancelled out if equivalent to an identity gate. This allows reducing the overall  circuit depth and thus the algorithm execution time. The fidelity can also be greatly improved as more operations can be executed within the qubit coherence time. \\


\begin{figure}[h]
\centering
\includegraphics[width=0.5\textwidth]{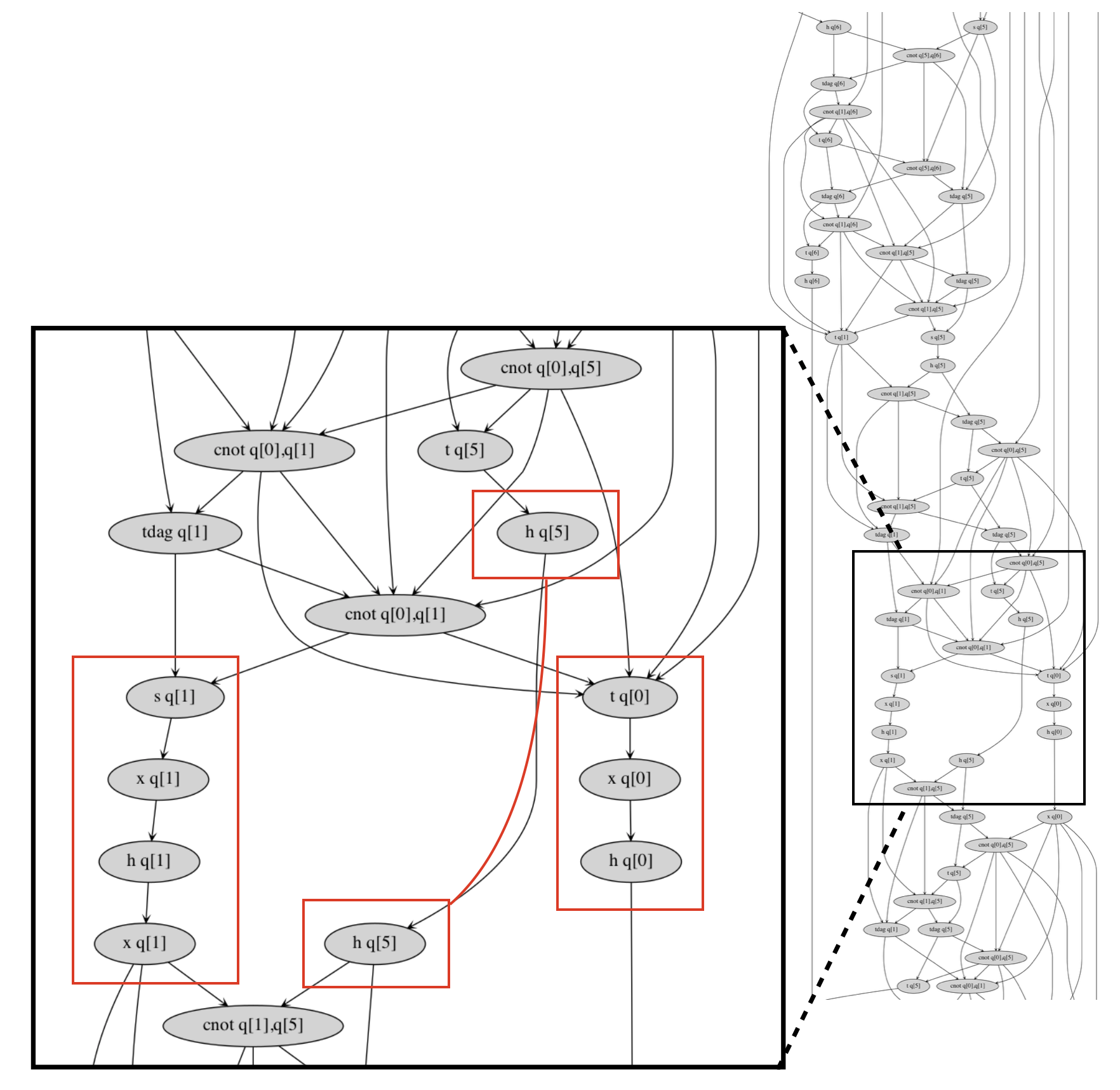}
\caption{Local optimization in gate-dependency graph: local sequences of single qubit operations can be merged into smaller sequence of elementary operations or cancelled-out when equivalent to an identity gate.}
\label{fig:local_optimization}
\end{figure}

Figure~\ref{fig:local_optimization} shows the gate dependency graph of a quantum circuit and a potential gate sequence optimization. We note, that without gate dependency analysis, some optimization opportunities can be missed as those gate sequences may be split into small scattered chunks that are not necessarily specified back-to-back in the original algorithm. \\

\subsubsection{Gate Sequence Optimization}

Gate sequence optimization uses the unitary representation of quantum gates to approximate the overall unitary operation. For instance, the equivalent unitary operation of a sequence of quantum gates operating on the same qubit can be obtained through matrix multiplication. The equivalent operation could be i) an identity that can be compiled out from the circuit, ii) an operation that can be implemented using a shorter sequence of elementary gates, iii) an operation that can be approximated using a shorter sequence of elementary operations. In order to control the accuracy of the compilation process, the compiler computes the distance between the target sequence of operation and the new set of elementary operations. The optimization will take place if that distance is smaller than the allowed error which is specified as a compilation parameter that can be controlled by the user to achieve at the desired accuracy.



OpenQL uses a sliding window over each sequence of gates to fuse locally quantum operations whenever possible. The size of the sliding window is critical to the compilation complexity which grows linearly with the number of gates. \\

\subsubsection{Gate Scheduling}

Gate scheduling aims to use gate-dependency analysis to extract parallelism and schedule the operations in parallel while respecting dependencies. It uses the knowledge of the duration of each gate as specified in the platform's configuration file to determine the cycle at which each gate can potentially start its execution.

OpenQL gate scheduling can perform three types of scheduling: an ASAP (As Soon As Possible), an ALAP (As Late As Possible) or a Uniform ALAP.

\begin{itemize}
    \item 
In an ASAP schedule, the cycle values are minimal but it may result in many gates being executed at the start of the circuit and thus longer cycles between successive gates operating on the same qubit, and thus a lower fidelity.
\item
At the other extreme, in an ALAP schedule the cycle values are maximal under the constraint that the total execution time of the circuit is equal to that of an ASAP schedule of the same circuit. But while at the start of the circuit relatively few gates are executed per cycle, at the end many gates will get executed on average. That they are executed as late as possible is good to get a higher fidelity but executing many gates per cycle may be more than the control electronics of the quantum computer was designed for, potentially leading to buffer overflows in that area and therefore to the requirement of a local feedback system to hold more gates off, effectively making execution time of a circuit longer.
\item
The Uniform ALAP schedule aims to produce an ALAP schedule with a balanced number of gates per cycle over the whole execution of the circuit. This scheduling scheme is based on \cite{BalancedSch}. It starts by creating an ASAP schedule then performing a backward pass over the circuit in an ALAP fashion: filling cycles with gates by moving them towards the end while respecting the dependencies.
\end{itemize}

\begin{figure}[h]
\centering
\includegraphics[width=0.45\textwidth]{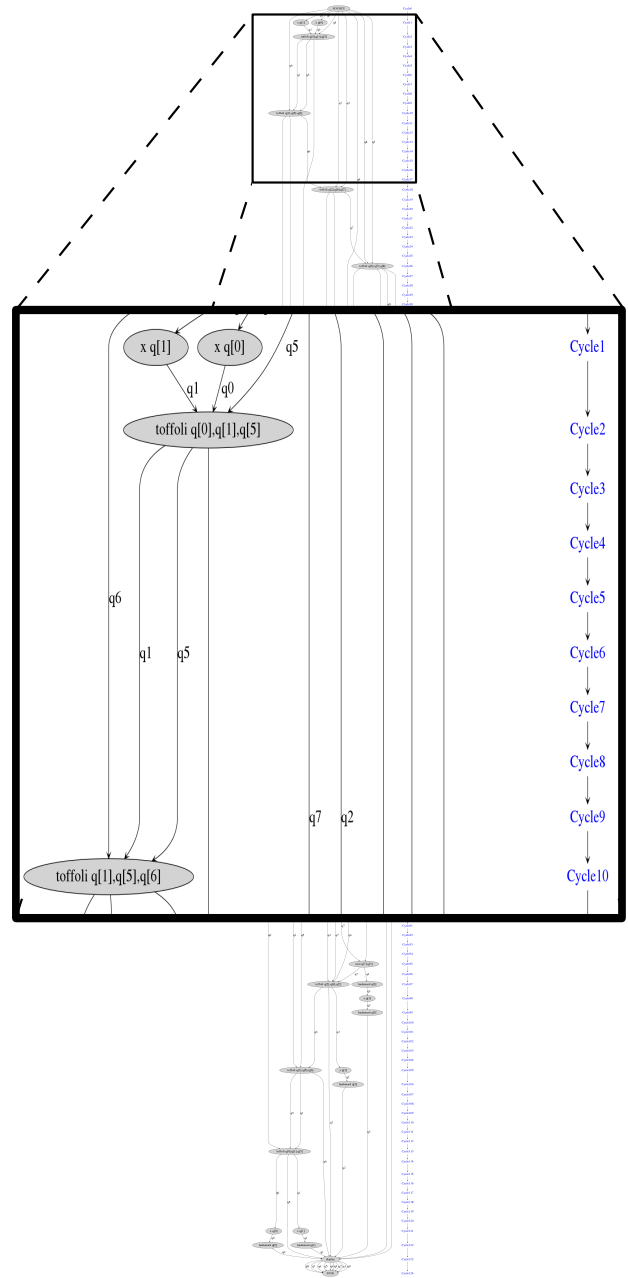}
\caption{Example of a As Soon As Possible (ASAP) Scheduling of the 3 Qubit Grover Algorithm.}
\label{fig:grover-schedule}
\end{figure}

Each of these three types of schedulers, dependencies and gate duration primarily determine the result. However, scheduler may need to respect more constraints, especially for the real targets. These constraints are mainly hardware constraints, for example those of control electronics, that limit the parallelism \cite{lao2019mapping}.


Using resource descriptions of those control electronics in the hardware configuration file, the gate scheduler optionally produces an ASAP, an ALAP or a Uniform ALAP schedule which respects these resource constraints. The main and from a hardware design perspective crucial property of the resulting schedules is that hardware can execute gates in the cycles determined by the scheduler as in a Very Long Instruction Word (VLIW) processor, without the need of maintaining whether gates are ready, etc.; this significantly reduces the complexity and size of the hardware.

\subsection{Mapping of quantum circuits}

The OpenQL compiler also includes the Qmap mapper~\cite{lao2019mapping} that is responsible for creating a version of the circuit that respects the processor contraints. The main constrains include the elementary gate set, the qubit topology that usually limits the interaction between qubits to only nearest-neighbour (NN) and the control electronics contraints -e.g. a single Arbitrary Waveform Generator (AWG) is used to operate in a group of qubits.

In order to adapt the circuit to these quantum hardware characteristics, the Qmap mapper: i) performs an initial placement of the qubits in which virtual qubits (qubits in the circuit) are mapped to the hardware qubits (physical qubits in the chip); ii) it will move non-neighbouring qubits to adjacent positions to perform a two-qubit gate; and iii) it will re-schedule the quantum operations respecting their dependencies and all hardware constraints. Note that it uses the hardware properties that are described in the configuration file.


The mapper aims to find the best qubit placement. Ideally, qubits can be placed in a way that all two-qubit interactions (two-qubit gates) present in the quantum program are allowed without need of any movement. However, this is rarely the case when the program is designed without considering the placement beforehand. Often qubit routing  is required to perform two-qubits operations between non-neighbouring qubits when the optimal placement does not allow direct interaction between them. From this perspective, qubit routing can be considered as a critical component of the qubit mapping which allow to resolve such conflicts.\\
OpenQL supports this by two algorithms, in sequence:
\begin{itemize}
    \item Initial Placement: This first pass aims to find the optimal qubit placement in the target physical device to enable performing two-qubits operations at the lowest possible cost. Currently, OpenQL can detect where constraints violations and thus illegal operations on such two-qubit gates between non-neighbouring qubits appear. It tries to find a map of the qubits that minimizes the overhead and enables qubit interactions. The mapper does this by using an Interger Linear Programming (ILP) algorithm as explained in \cite{lao2018mapping}. Such an approach works perfectly on smaller circuits but takes too much execution time on longer circuits because of exponential scaling.
    \item Qubit router: The second pass guarantees that two-qubit gate operations on non-neighbouring qubits can be performed by inserting a series of gates -e.g. SWAP gates- that move qubits to neighbouring places. For each of such two-qubit gate operations, it determines the distance of those qubits and when too far apart, it evaluates all possible ways to make those qubits nearest neighbour. To do so, it  evaluates all possible shortest paths and chooses the one that, for instance, results in the minimum increase of the circuit depth (number of cycles). Then, the corresponding 'move' operations are inserted in the program.

\end{itemize}
Note that after the mapping the number of gates and the circuit depth will increase, increasing the failure rate and then reducing the algorithm's reliability.

\subsection{Technology-Independent Common QASM}



After gate decomposition, quantum circuit optimization or gate scheduling, a cQASM compiler is responsible of producing a technology-independent common quantum assembly code called cQASM. Currently the cQASM 1.0 \cite{khammassi2018cqasm} is used to describe the circuit at the gate level and allows the user to simulate the execution of the quantum algorithm using the QX Simulator \cite{Khammassi2017}. The simulation allows the programmer to verify the correctness of the quantum algorithm or to simulate and evaluate its behaviour on noisy quantum computing devices. \\

The cQASM 1.0 aims to enable the description of quantum circuit while abstracting away the hardware details, for instance, a \verb|H q[1]| describes a Hadamard gate on qubit q[1] without specifying the low level implementation of that quantum operation on a specific qubit technology. Besides the description of common quantum operations, cQASM 1.0 allows the specification of parallelism in the quantum circuit in the form of 'bundles' (lists of gates starting in a same cycle) and 'SIMD operations' (a gate operating on a range of qubits). This allows the OpenQL scheduler to express the parallelism that it found in cQASM 1.0.\\

The cQASM 1.0 allows the naming of quantum circuit sections or "sub-circuits"; these sub-circuits correspond to the names of the quantum kernels and allow the user to relate the produced cQASM to its high-level algorithm written in Python or C++. \\  

In the cQASM code example \ref{code:Grover}, we see the scheduled code produced for the Grover search algorithm. 

\begin{lstlisting}[
caption=Grover Algorithm.,
label=code:Grover,
float,floatplacement=H, 
belowskip=-1.0 \baselineskip
]
   version 1.0
    
   # define a quantum register of 9 qubits
   qubits 9
 
    # sub-circuit for state initialization
    .init
        x q[4]    # oracle qubit
        h q[0:4]  # parallel hadamard gates on qubits 0,1,2,3 and 4
   
    # core step of Grover's algorithm
    # loop with 3 iterations
    .grover(3)

       # search for |x> = |0100>
       
       # oracle implementation
       x q[2] 
       toffoli q[0],q[1],q[5]
       toffoli q[1],q[5],q[6]
       toffoli q[2],q[6],q[7]
       toffoli q[3],q[7],q[8]
       cnot q[8],q[4]
       toffoli q[3],q[7],q[8]
       toffoli q[2],q[6],q[7]
       toffoli q[1],q[5],q[6]
       toffoli q[0],q[1],q[5]
       x q[2]

       # Grover diffusion operator
       { h q[0] | h q[1] | h q[2] | h q[3] } # parallel gates
       { x q[0] | x q[1] | x q[2] | x q[3] }
       h q[3]
       toffoli q[0],q[1],q[5]
       toffoli q[1],q[5],q[6]
       toffoli q[2],q[6],q[7]
       cnot q[7],q[3]
       toffoli q[2],q[6],q[7]
       toffoli q[1],q[5],q[6]
       toffoli q[0],q[1],q[5]
       h q[3]
       { x q[0] | x q[1] | x q[2] | x q[3] }
       { h q[0] | h q[1] | h q[2] | h q[3] }
       display

   # final measurement
   .measure
       h q[4]
       measure q[4]
       display
\end{lstlisting}

    \subsection{Technology-Dependent Compilation : eQASM}
\label{sec:eqasm}
After compiling the technology-independent QASM code, the compiler generates the Executable QASM (eQASM) which targets specific control hardware. The compiler uses different eQASM compilation backends depending on the target platform specified in the hardware configuration file. The eQASM compiler can reschedule the quantum operations to exploit the available parallelism on the target micro-architecture and map the quantum circuit based on the topology of the target qubit chip and the connectivity of the control hardware.

\subsection{Quantum Computer Micro-Architecture}
OpenQL has currently several backends capable of generating executable Quantum Assembly Code (eQASM) for two different microarchitectures discussed in \cite{Fu2017} and \cite{eQASM}. The backends convert the compiled cQASM code to a specific eQASM code for the target microarchitecture with respect to the hardware constraints such as the available parallelism and the timing constraints. 

\subsubsection{Temporal Transformation : Low-level Scheduling}
While the QASM-level scheduler pass extracts all the available gate-level parallelism, the target platform can have limited parallelism due the control electronic constraints. After analyzing the quantum gate dependencies, the compiler schedules the instructions either ‘As Late As Possible’ (ALAP) or ‘As Soon As Possible’ (ASAP) with respect to the gate dependencies and cycle-accurate durations of the different gates.

\subsubsection{Spatial Transformation : Connectivity-Aware Mapping}
The OpenQL compiler maps the qubits with respect to the qubit plane topology which specifies the operation constraints such as nearest neighbour interactions or operation parallelism limitations. The current version of OpenQL relies on the two-qubit instruction specification in the hardware configuration file to extract the constraints, but the mapping task is being shifted to the mapping layer at the gate level which will use a dedicated mapping specification in the hardware configuration file and more advanced mapping techniques.


\subsubsection{eQASM Execution Monitoring}
\begin{figure*}[h]
\centering
\includegraphics[width=0.75\textwidth]{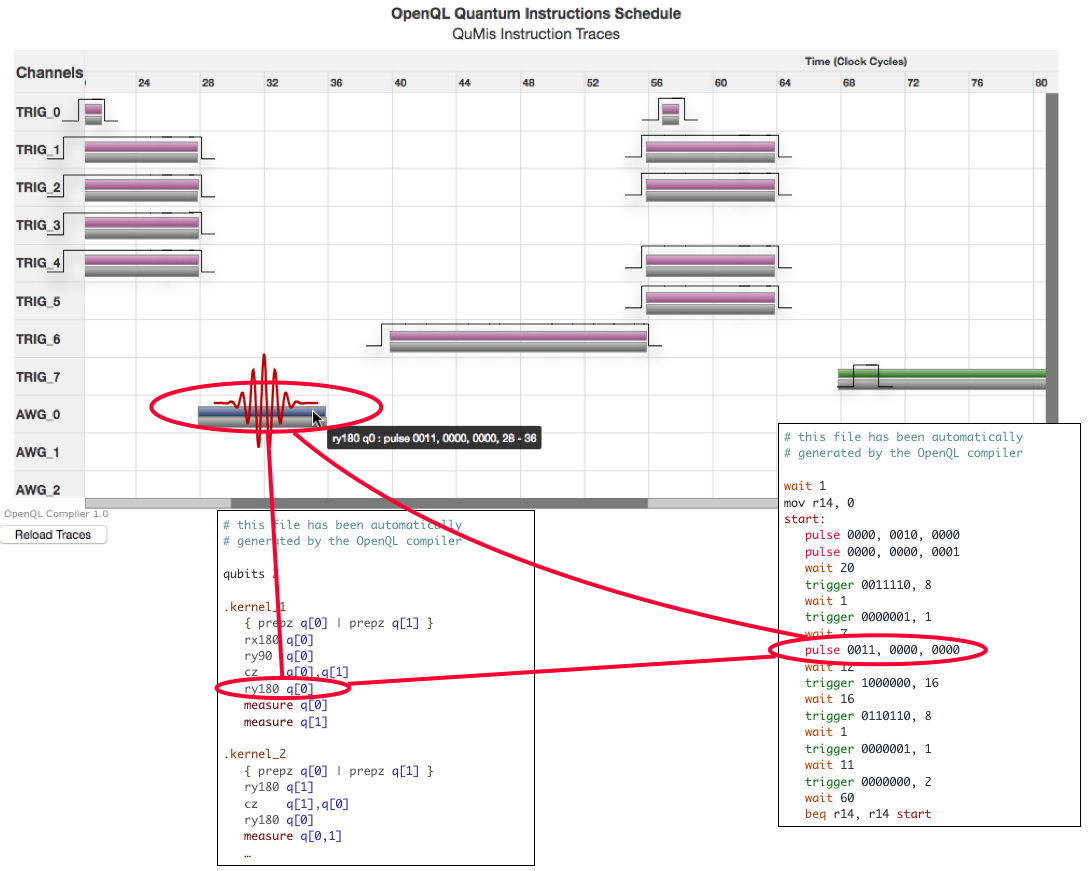}
\caption{Instruction Timing Diagram generated by OpenQL}
\label{fig:instruction_tracing}
\end{figure*}

Tracing the different the instruction execution and timing of the different signals controlling the qubits is critical for debugging and monitoring the hardware. The OpenQL compiler generates auxilliary outputs for tracing purposes such as timed instructions and a graphical timing diagram as shown in Fig. \ref{fig:instruction_tracing}. In this timing diagram, both the digital and analog signals are shown with their respective starting time and duration. Each signal refer to both its originating eQASM instruction and the originating cQASM instruction with the precise execution clock cycle. When the compiler compensate for latencies in a given channel, both the original and the compensated timing are shown.   


    \subsection{Hardware Configuration Specification : Control Electronics }
\label{subsec:hw_config}
In order to compile the produced QASM instructions into executable instructions (e.g. eQASM), the compiler needs to know not only the instruction set supported by the target microarchitecture but also the specification of all the constraints related the hardware resource usage, the operations timing and the qubits connectivity, etc. \\

The hardware specification file aims to provide this information in an abstract way to allow describing different architectures and enable the compiler to adapt to their constraints and requirements when producing the executable code. This allows extending the compiler support to many architectures without fundamental changes in its upper technology-independent layers. \\

The following JSON code describe the hardware setup and list all the supported operation and their settings such as the number of qubits, the time scale, the operations dependencies, their timing parameters, mathematical description and associated instruction set. 

\begin{lstlisting}[language=json][caption=OpenQL Hardware Configuration File]
{
   "eqasm_compiler" : "qumis_compiler",

   "hardware_settings": {
         "qubit_number": 2,
         "cycle_time" : 5,
         "mw_mw_buffer": 0,
         "mw_flux_buffer": 0,
         …
   },
   "instructions": {
      "rx180 q1" : {
         "duration": 40,
         "latency": 20,
         "qubits": ["q1"],
         "matrix" : [ [0.0,0.0], [1.0,0.0],
                      [1.0,0.0], [0.0,0.0] ],
         "disable_optimization": false,
         "type" : "mw",
         "qumis_instr": "pulse",
         "qumis_instr_kw": {
            "codeword": 1,
            "awg_nr": 2
         }
      },
      "rx180 q0" : {
         "duration": 40,
         "latency": 10,
         "qubits": ["q0"],
         "matrix" : [ [0.0,0.0], [1.0,0.0],
                      [1.0,0.0], [0.0,0.0] ],
         "disable_optimization": false,
         "type" : "mw",
         "qumis_instr": "codeword_trigger",
         "qumis_instr_kw": {
            "codeword_ready_bit": 0,
            "codeword_ready_bit_duration" : 5,
            "codeword_bits": [1, 2, 3, 4],
            "codeword": 1
         }
     },
     "prepz q0" : {
         "duration": 100,
         "latency": 0,
         "qubits": ["q0"],
         "matrix" : [ [1.0,0.0], [0.0,0.0],
                    [0.0,0.0], [1.0,0.0] ],
         "disable_optimization": true,
         "type" : "mw",
         "qumis_instr": "trigger_sequence",
         "qumis_instr_kw": {
            "trigger_channel": 4,
            "trigger_width": 0
         }
      }
   },

   "gate_decomposition": {
      "x q0" : ["rx180 q0"],
      "y q0" : ["ry180 q0"],
      "z q0" : ["ry180 q0","rx180 q0"],
      "h q0" : ["ry90 q0"],
      "cnot q0,q1" : ["ry90 q1","cz q0,q1","ry90 q1"]
   },

   "resources" : {
   },

   "topology" : {
   }
     …
}
\end{lstlisting}

The sections of the hardware configuration file are organized as follows:
\begin{itemize}
    \item \textbf{eqasm\_compiler}: this section specifies the executable QASM (eQASM) compiler backend which should be used to generate the executable code. The allows the compiler to target different microarchitectures using the appropriate backend.
    \item \textbf{instructions}: in this section, the quantum operations supported by the target platform are described by their duration, their latency in the control system, their unitary matrix representation, their type (microwave, flux or readout) and finally microarchitecture-specific information to enable the compiler to generate the executable code.
    \begin{itemize}
        \item \textbf{Instruction Properties}
        \begin{itemize}
            \item \textit{duration} (int) : duration of the operation in ns
            \item \textit{latency} (int): latency of operation in ns
            \item \textit{qubits} (list) : list of affected qubits by this operation (this includes the qubits which are directly used or made inaccessible by this operation).
            \item \textit{matrix} (matrix): the unitary matrix representation of the quantum operation.
            \item \textit{disable\_optimization} (bool): setting this field to True prevent the compiler from compiling away or optimizing the operation.
            \item \textit{type} (str): one of either 'mw' (microwave), 'flux' , 'readout' or 'none'.
        \end{itemize}
        \item \textbf{Microarchitecture Specific Properties}
        \begin{itemize}
            \item \textit{qumis\_instr} (str): one of wait, pulse, trigger, CW\_trigger, dummy, measure.
            \item \textit{qumis\_instr\_kw} (dict): dictionary containing keyword arguments for the qumis instruction.
        \end{itemize}
    \end{itemize}
    \item \textbf{gate\_decomposition}: the gate decomposition section aims to describe the decomposition of  coarse grain quantum operations into the elementary operations defined in the previous section. Each composite instruction in this section is defined by its equivalent quantum gate sequence.
    For instance, a CNOT gate can be described as:
    "CNOT 
    \item \textbf{resources}: describe the various hardware constraints that are used by the hardware constrained scheduling algorithm
    \item \textbf{topology}: describes the qubit grid topology, i.e. qubits and their connnections for performing two-qubit gates
\end{itemize}

The operation duration, latency and the target qubits are used by the eQASM backend to analyze the dependencies of the instructions. This information is critical for different compilation stages, for instance the duration of an instruction and its qubit dependency is crucial for the low-level hardware-dependent scheduling stage which use these information to schedule the instructions. \\

The latency field is used by the backend compiler to compensate for the instruction latency by adjusting the instructions starting times to synchronize different channels with different latencies. Different latencies could exist in different control channels due to propagation delays through different cables, control latencies in waveform generators or readout hardware.
\section{OpenQL Application}
\label{sec:results}


OpenQL has been used to program several experiments and algorithms on various quantum computer architectures and also on different qubit technologies, namely superconducting and semiconducting qubits. 


\subsection{Superconducting Qubit Experiments}

We used OpenQL to compile quantum code and implement various experiments on several quantum chips with 2, 5 and 7 qubits using two different microarchitectures, namely QuMA 1.0 \cite{Fu2017} and QuMA 2.0 \cite{Fu2018}, for controlling the qubits using two different instruction sets.
We implemented several standard experiments such as Clifford-based Randomized Benchmarking \textit{RB}~\cite{magesan2011scalable}, \textit{AllXY}~\cite{reed2013entanglement} and other calibration routines, such as Rabi oscillation~\cite{reed2013entanglement}. For each experiment, the same high-level OpenQL code has been reused on different setups and devices without changes, only the hardware configuration file has been changed to specify each target hardware setup and its constraints to instruct the compiler how to generate the appropriate code for each platform. Apart from the above basic experiments, OpenQL has also been used to compile code for the following applications:

\begin{enumerate}
    \item Net-zero two qubit gate~\cite{Rol19_NetZero}
    \item 3 qubit repeated parity checks~\cite{Bultink19_ZZXX} 
    \item Variational quantum eigen solver~\cite{Sagastizabal19}
    \item Calculating energy derivatives in quantum chemistry~\cite{OBrien19} 
\end{enumerate}{}

\subsection{Semiconducting Qubit}

In order to evaluate the portability of OpenQL over different qubit technologies, the AllXY experiment has been reproduced on both superconducting qubit and semiconducting qubit using the same code and different configuration files. We used a Si-Spin qubit device \cite{watson2018} controlled by different control electronics, the hardware configuration file was changed to reflect the control setup and enable the compiler to automatically adapt the generated code to the target system: the compiler took into account the latencies of the different signal generators and measurement units involved in the setup and rescheduled all the quantum operations accordingly to compensate for those latencies and provide coherent qubit control. 

\section{Conclusion}
\label{sec:conclusion}

In this paper we presented the OpenQL quantum programming framework which includes a high-level quantum programming language and its compiler. A quantum program can be expressed using C++ or Python interface and compiler translates this high-level program into a Common QASM (cQASM) to target simulators. This program can further be compiled for a specific architecture targeting physical quantum computer. OpenQL has been used for implementing several experiments and quantum algorithms on several quantum computer architectures targeting both superconducting and semiconducting qubit technologies.

\section*{Acknowledgment}
This project is funded by Intel Corporation. The authors would like to thank Pr. L. DiCarlo and his team for giving us the opportunity to test OpenQL on various Superconducting qubit systems and Pr. L. Vandersypen and his team for allowing us to test OpenQL on a Si-Spin qubit device. The authors would like to thank all members of the Quantum Computer Architecture Lab at TU Delft for their valuable feedback and suggestions.

\bibliographystyle{IEEEtran}
\bibliography{references}

\end{document}